\documentclass[aps,prb,preprint,superscriptaddress]{revtex4-1}

\usepackage{graphicx}
\usepackage{epstopdf}
\usepackage{float}
\begin{document}


\title{Optical Response of DyN}


\author{M. Azeem}\author {B. J. Ruck}
\affiliation{The MacDiarmid Institute for Advanced Materials and Nanotechnology, School of Chemical and Physical Sciences, Victoria University,
P.O. Box 600, Wellington 6140, New Zealand}
 \author{ Binh Do Le} \author{H. Warring} 
\affiliation{The MacDiarmid Institute for Advanced Materials and Nanotechnology, School of Chemical and Physical Sciences, Victoria University,
P.O. Box 600, Wellington 6140, New Zealand}

\author{ N. M. Strickland} \author { A. Koo }
\affiliation{Industrial Research Limited, Lower Hutt, P.O. Box 31310, Lower Hutt 5040, New Zealand}

\author {V. Goian}
\author {S. Kamba}
\affiliation{Institute of Physics, Academy of Sciences of the Czech Republic, Na Slovance 2, 182 21 Prague 8, Czech Republic}

 \author {H. J. Trodahl }
\affiliation{The MacDiarmid Institute for Advanced Materials and Nanotechnology, School of Chemical and Physical Sciences, Victoria University,
P.O. Box 600, Wellington 6140, New Zealand}




\begin{abstract}
We report measurements of the optical response of polycrystalline DyN thin films. The frequency-dependent complex refractive index in the near IR-visible-near UV was determined by fitting reflection/transmission spectra. In conjunction with resistivity measurements these identify DyN as a semiconductor with 1.2~eV optical gap. When doped by nitrogen vacancies it shows free carrier absorption and a blue-shifted gap associated with the Moss-Burstein effect. The refractive index of $2.0 \pm 0.1$ depends only weakly on energy. Far infrared reflectivity data show a polar phonon of frequency  280~cm$^{-1}$ and dielectric strength $\Delta\epsilon= 20$. 
\end{abstract}


\maketitle
\section{Introduction}
Nitride compounds of rare-earth (RE) ions have gained attention due to their interesting magnetic and electronic properties. With the exception of Ce the RE ions are in their preferred trivalent state in the nitrides, so that their magnetic characters originate from their incompletely filled \textit{4f} shell. They are predicted to be half metals or ferromagnetic semiconductors; thus they are strong candidates for use in spintronic devices.\cite {Ruck08, Ivanov04, larson07,aerts04,granville09}\

Among the rare-earth nitride (REN) family, GdN is the most thoroughly studied\cite {granville06, leu05, khazen06, mitra08, sharma10} compound. The Gd$^{3+}$  ion has half filled 4\textit{f} shell with spin moment\cite {ludbrook09} of 7$\mu_B$. Its nitride is now a well established ferromagnetic semiconductor \cite {granville06} with T$_C$=70~K, though a lower \cite {yosh11, yoshprb} T$_C$=30 K is also reported. It has an optical energy gap  \cite {yosh11,yoshprb, tro07} of 1.3 eV in its paramagnetic state and 0.98 eV in ferromagnetic state. Most of the other RENs are known to be ferromagnetic with lower Curie temperatures; the exception\cite {rich11} is EuN , which cannot order due to the $J=0$ state of the 4\textit{f} shell in Eu$^{3+}$ .  

The present work describes an experimental study of DyN. It is a ferromagnetic semiconductor with a reported\cite {preston07, pokrant01} T$_C$ ranging between 17 and 26 K.  LSDA+U electronic structure calculations \cite {larson07} predict that seven spin-majority 4\textit{f} states occur in three deep narrow bands while two 4\textit{f} electrons go in minority-spin bands 5~eV below the top of the valence band. The same study shows a small indirect gap between the top of the valence band at $\Gamma$ and the conduction band minimum at X and a minimum  direct gap of 1.17~eV at X. 

There are decades-old reports \cite {sclar64, bush69} of absorption edges of almost all RENs, with the exception of CeN and PmN, though it is now recognised that those early materials were subject to the formation of nitrogen vacancies and decomposition as oxides in air.  In particular DyN has been reported to show an onset of  absorption ranging\cite {sclar64, bush69, dismukes70}  from 2.9~eV to 0.91~eV.  Preston \cite {preston07} \textit{et al.} reported the measured gap value of $\sim$ 1.5 eV between x-ray absorption and emission spectroscopies. In the wake of these widely deviating claims, there is a vital need for a systematic experimental study of DyN. Here we report reflection/transmission measurements from 0.5~eV to 5.0~eV to determine the optical gap and seek evidence of optical transitions above gap, and far infrared reflection measurements of the zone-center phonon frequency.


\section{Experimental Details}

Thin films of DyN were prepared at ambient temperature by depositing Dy at a rate of 0.5-2~\AA s$^{-1}$ in the presence of 10$^{-4}$ - 10$^{-5} $ mbar  of carefully purified N$_2$, as has been described in more detail previously.\cite {granville06} It is expected that, as is true for GdN, a high concentration of nitrogen-vacancy donors will be found in any but the films grown in the highest N$_2$ pressures. The nitrogen vacancies each bind either one or two electrons, leaving at least one electron in the conduction band.\cite{punya11}

 Before depositing the films the chamber was evacuated to a base pressure of less than 10$^{-8}$ mbar, and residual gasses were reduced further during deposition by the gettering effect of Dy.  Due to the propensity of the rare-earth nitride thin films to atmosphere, these films need to be passivated with a capping layer. The choice of substrate and cap was dictated by the measurements to be made: sapphire and MgF$_2$ for the near-visible range, yttria-stabilised zirconia and Si for  the far infrared. XRD was performed to establish the crystal structure, lattice constant and orientation of the films.

 Magnetic measurements were performed with a Quantum Design MPMS superconducting quantum interference device. The resistances of the films were monitored both \textit{in situ} during growth and \textit {ex situ} as a function of temperature.

Transmission and reflection spectra were obtained for Al$_2$O$_3$/DyN/MgF$_2$ in the energy range of 0.5-2.0 eV using a Fourier transform spectrometer (BOMEM model DA8) and from 1 to 6 eV using a conventional visible-UV spectrometer.  A gold film and quartz wedge were used as the comparison standard for reflectance measurements in the infra-red and visible regions respectively. Reflectance measurements were performed for light incident on both the film and substrate surfaces,  but since the transmittance is unaffected by the direction that light traverses through the sample it was taken from one side alone. The partially reflected and transmitted rays interfere to form a complex interference pattern that can compete with the loss of transmission signalling the absorption edge. In order to extract the optical constants of the DyN layer a commercial software TFCalc was used which makes use of characteristic matrix method. 

The unpolarized near-normal infrared (IR) reflectance spectra were taken using a Bruker IFS 113v FTIR spectrometer in the spectral range of 30-3000 cm$^{-1}$ with a resolution of 2 cm$^{-1}$. Each of the reflectance spectra was evaluated as a two-layer optical system. \cite{zele98} At first, the bare substrate reflectivity was measured and carefully fitted using the generalized factorized damped harmonic oscillator model

\begin{equation}
{\epsilon^*}(\omega)={\epsilon}' - i{\epsilon}''=\epsilon_\infty \prod _j \frac{\omega_{LOj}^{2}-\omega^2+i\omega\gamma_{LOj}}{\omega^{2}_{TOj}-\omega^2+i\omega\gamma_{TOj}} ,
\end{equation}

\noindent where $\omega_{LOj}$   and  $\omega_{TOj}$ are transverse and longitudinal frequencies of the $j$-th polar phonon, respectively,  $\gamma_{LOj}$ and $\gamma_{TOj}$   are their damping constants, and $\epsilon_\infty$  denotes the high frequency permittivity resulting from electronic absorption processes. The complex dielectric function $\epsilon^*(\omega)$ is related to the reflectivity $R(\omega)$ of the bulk substrate by 

\begin{equation}
R(\omega)= \left |\frac{\sqrt{\epsilon^*(\omega)}-1}{\sqrt{\epsilon^*(\omega))}+1}  \right |^2.
\end{equation}

The high-frequency permittivity $\epsilon_\infty= 5.88$ of the substrate resulting from the electronic absorption processes was obtained from the frequency independent reflectivity tail above the phonon frequency. When analyzing the reflectance of the substrate together with the film, we used the bare substrate parameters and adjusted only the dielectric function of the film. For this purpose, we preferentially used a classical three-parameter damped oscillator model

\begin{equation}
\epsilon^*(\omega)=\epsilon_\infty+\sum_{j=1}^{n} \frac{\Delta\epsilon_j\omega^2_{TOj}}{\omega^{2}_{TOj}-\omega^2+\iota\omega\gamma_{TOj}},
\end{equation}

\noindent where $\Delta\epsilon_j$ is the dielectric strength of the $j$-th mode.

\section {Results and Discussions} 

\begin{figure} [H]
\centering
\includegraphics [scale=0.50]{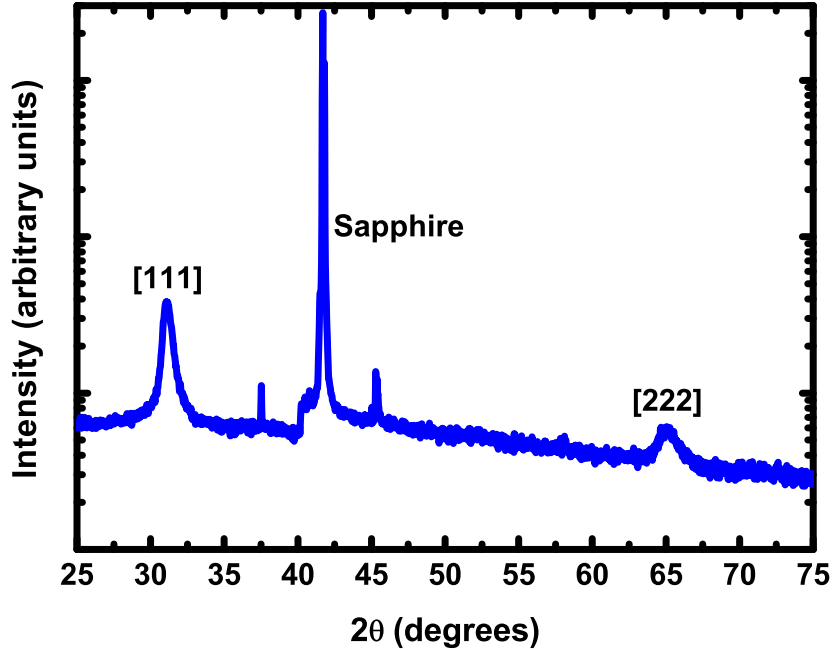}
\caption{(Color online) XRD pattern for a representative DyN thin film. The most prominent peak comes from the sapphire substrate. Peaks labelled [111] and [222] are contributed by strongly textured DyN. }
\label{xrd}
\end{figure}

Figure~\ref{xrd} shows the XRD scan of a typical DyN film, in this case on sapphire and with a capping layer of MgF$_2$. The strongest peak comes from the sapphire substrate while the next prominent peak labelled as [111] and a rather weak [222] peak are attributed to the cubic structure of DyN. The films are strongly [111] textured, similar to other RE nitrides grown at ambient temperature \cite {granville06, preston07}. The lattice constant of the films is  0.490~nm as expected\cite{preston07} and the average crystallite size is about 10 nm as obtained using Scherrer formula. There are no secondary phases detected in the XRD spectra. 

\begin{figure} [h]
\centering
\includegraphics [scale=0.50]{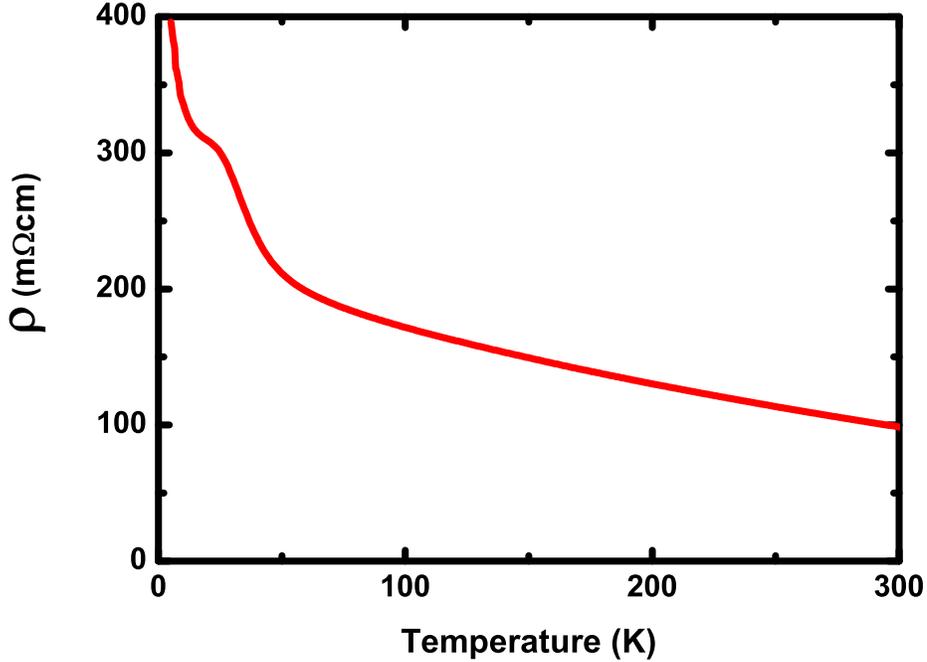}
\caption{(Color online) Temperature dependent resistivity of a DyN thin film establishing the semiconducting nature of DyN.}
\label{resis}
\end{figure}

Figure~\ref{resis} shows the  the temperature-dependent resistivity of a film grown at high N$_2$ pressure. The room-temperature resistivity of 100 m$\Omega$ cm leads then to a carrier concentration of less than 10$^{20}$ cm$^{-3}$, characteristic of a moderately doped semiconductor for assumed mean free paths of 1-10 nm. The semiconducting nature of the film is  confirmed by a strongly rising resistivity with decreasing temperature.  A relatively flat peak near the ferromagnetic Curie temperature ($T_C$, see below) is then followed at lower temperature by a continuation of the rise, affirming a semiconducting ground state below $T_C$.  

The magnetic susceptibility follows the Curie-Weiss expectation with an estimated Curie temperature of 20 K. However, the hysteretic behaviour of the lower-temperature ferromagnetic phase persists to higher temperature so we quote T$_C$ as lying between 20 K and 25 K, in agreement with the higher values found in the literature.\cite{preston07} 

\begin{figure} 
\centering
\includegraphics [scale=0.50]{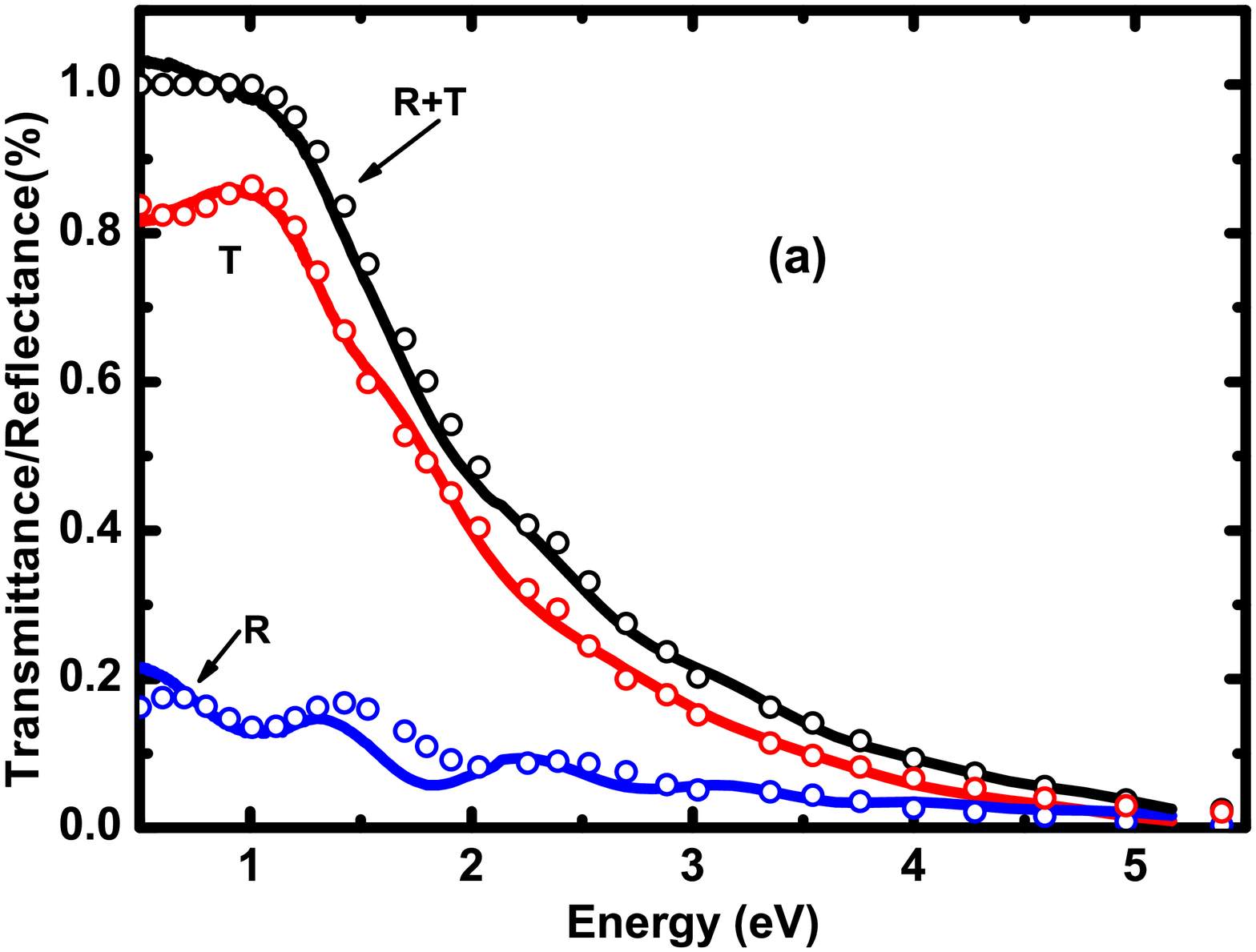}
\includegraphics [scale=0.50]{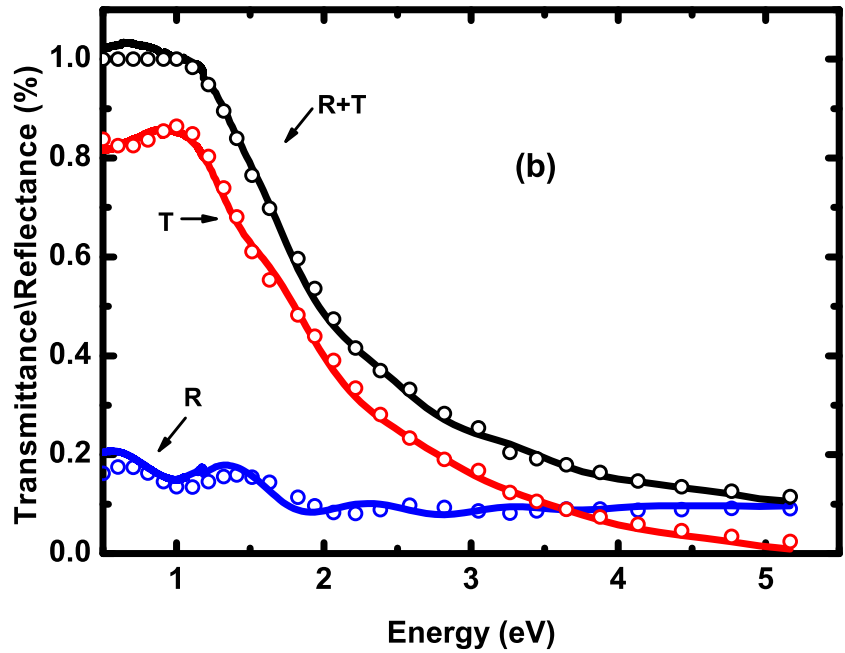}
\caption{ (a) (Color online) Reflectance from the cap side, transmittance and sum of R and T from $\approx$ 300~nm thick DyN film protected by a MgF$_2$ capping layer. Transmission drops after 1.2~eV indicating the presence of an optical gap. Solid lines are experimentally obtained spectra whereas open circles represent fitted spectra. (b) Optical spectra for the same film showing reflectance and transmittance from the substrate side. }
\label{RT}
\end{figure}

Turning our attention towards the main features of this work, Figure~\ref{RT} shows reflectance (R), transmittance (T) and their sum for a DyN film grown under a high N$_2$ partial pressure, as obtained from its cap side. Focusing on the low energy region (0.5 eV-1.0 eV) first, we find that the absorptance (1-R-T) is zero within 2$\%$  uncertainty, as  establishing a very low free carrier density expected of a semiconductor and signalling that this energy range is below the interband edge. Above 1.2~eV the transmitted light falls gradually indicating the presence of interband transitions. 

The interpretation of the R/T spectra was accomplished assuming the refractive indices for the MgF$_2$ cap and sapphire substrate as 1.4 and 1.8, respectively. To first approximation, the absorption below the edge was initially set to zero, as is in any case indicated by R+T=1. The refractive index of 2.0 was then determined by fitting the transmission spectra below the band edge; even the average value of transmission ensures that this is the refractive index in this energy range. Next, with this value of the refractive index approximated as constant above the edge, values of \textit{k} were extracted by fitting the absorption spectra. Spectral dependence of refractive index was then allowed, but the variations were below the level of confidence so we quote a refractive index of 2.0$\pm0.1$. The circles in Figure~\ref{RT} show a comparison between the calculated and measured R/T spectra. We regard the fit as reasonable; the computer program calculates optical spectra for perfect interfaces and uniform films, but in reality one expects the films to show some degree of interface roughness and also we do not know thickness of the film very accurately. 

\begin{figure} 
\centering
\includegraphics [scale=0.50]{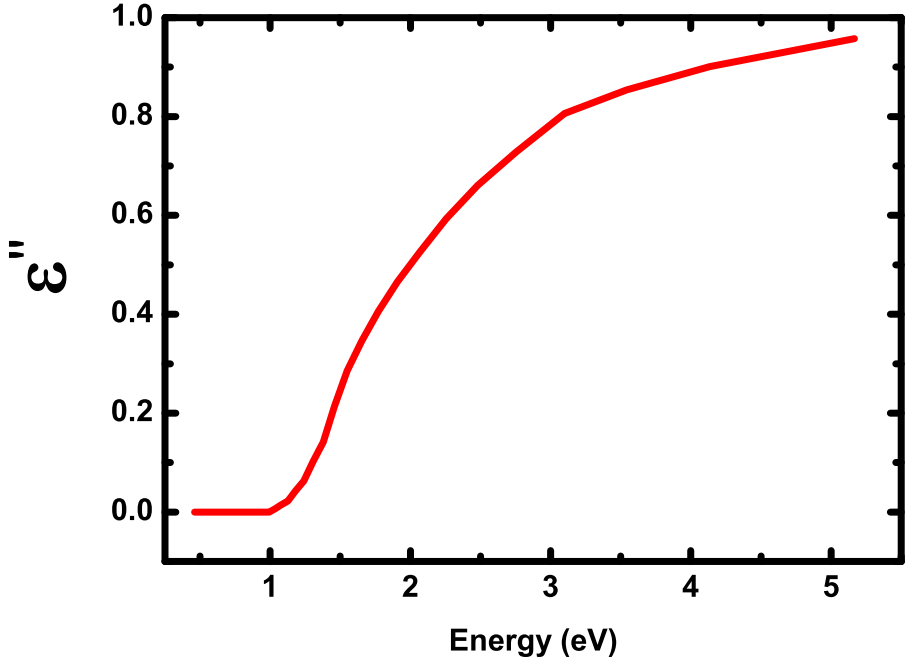}
\caption{(Color online) Imaginary part of dielectric function depicts the fundamental absorption edge at 1.2 eV for a near-stoichiometric DyN film. }
\label{epsi}
\end{figure}

Figure~\ref{epsi} shows the imaginary part of the dielectric function calculated by using $\epsilon^{\prime\prime}=2nk$ where \textit{n} and \textit{k} were obtained from the fits above. The absorption increases monotonically with energy, showing no structure that might result from interband onsets at any energy above the first optical absorption edge.  The rapid drop near the edge extrapolates to a gap of about 1.2~eV, with a tail to lower energy that we believe is related to uncertainties in the parameters due to incomplete correction for the interference fringes. It agrees within uncertainty with the X-point gap of 1.17 eV predicted by Larson\cite{larson07} \textit{et al.}  

Two further films have been studied, grown with substantially smaller excess nitrogen flux as indicated in Table I. As expected the higher density of N$_2$ vacancies in these films lead to free-carrier absorption below 1.2 eV, with absorption coefficient at 0.5 eV also listed in the Table.  It is clear that a reduced N$_2$/Dy ratio leads to sub-gap absorption, as is expected for the higher density of nitrogen vacancy dopants that has earlier been reported for a lowered N$_2$ pressure during rare earth nitride growth.\cite {granville06} 

Figure 5 illustrates the relation between free carriers and band gap with N$_2$/Dy flux ratio during growth. Film A, grown with a N$_2$/Dy ratio of 250, is close to stoichiometric, and accordingly the subgap absorption is below the measurement limit. Films B and C, grown with lower N$_2$/Dy flux ratio, have a larger concentration of N$_2$ vacancies and finite sub-gap absorption. To estimate the free-carrier concentration we note that in the high frequency limit $\omega\tau\gg1$ the absorption coefficient is given by 

\begin{equation}
\alpha = \frac{4\pi}{\lambda}\left (\frac{\sigma_{DC}}{2n\epsilon_0\omega^3\tau^2}  \right )
\end{equation}

Applying this to the films in question, and assuming an effective mass in the conduction band of m$^*\approx$~0.2 estimated from the DyN bandstructure,\cite {larson07} the concentration of free carriers in film A was estimated to be $<$10$^{20}$ cm$^{-3}$, in agreement with the inference drawn above from the resistivity.  For films grown at lower N$_2$/Dy flux we have found carrier concentrations of order 10$^{21}$  cm$^{-3}$.  Those carriers are accommodated in the three electron pockets at X, and then introduce a degenerate electron gas of Fermi energy 0.2 eV and 0.3 eV in films B and C, respectively, in agreement with the Moss-Burstein shift of the absorption edge seen in Fig.~5. 

\begin{center}
\begin {table} 
\caption {Growth parameters for various DyN thin films of approximately 300nm thickness. } \

\begin{tabular}{| c|c|c|c |c |c| } 
\hline
  & \ \ \ Growth Pressure \ \ &  \ \ Deposition Rate \ \ & \ \  N$_2$/Dy Flux Ratio \ \  &  \ \ $\alpha$ at 0.5 eV \ \ & \ \  Direct Gap \ \ \\
 & (mbar) & (nm/s) & & (10$^{3}$ cm$^{-1})$ & (eV)  \\ \hline
\ \ Film A \ \ & $1.3 \times 10^{-4}$ & $ 0.05 $& $250$ & 0& $1.2$  \\  \hline
\ \ Film B \ \ & $1.7 \times 10^{-4}$ & $ 0.15 $& $75$ & 6.5 &$1.5$  \\  \hline
\ \ Film C \ \ & $7.0 \times 10^{-5}$ & $ 0.2 $& $22$ & 9.7 &$1.7$  \\  \hline
\end{tabular}
\end{table}
\end {center}

\begin{figure} 
\centering
\includegraphics [scale=0.50]{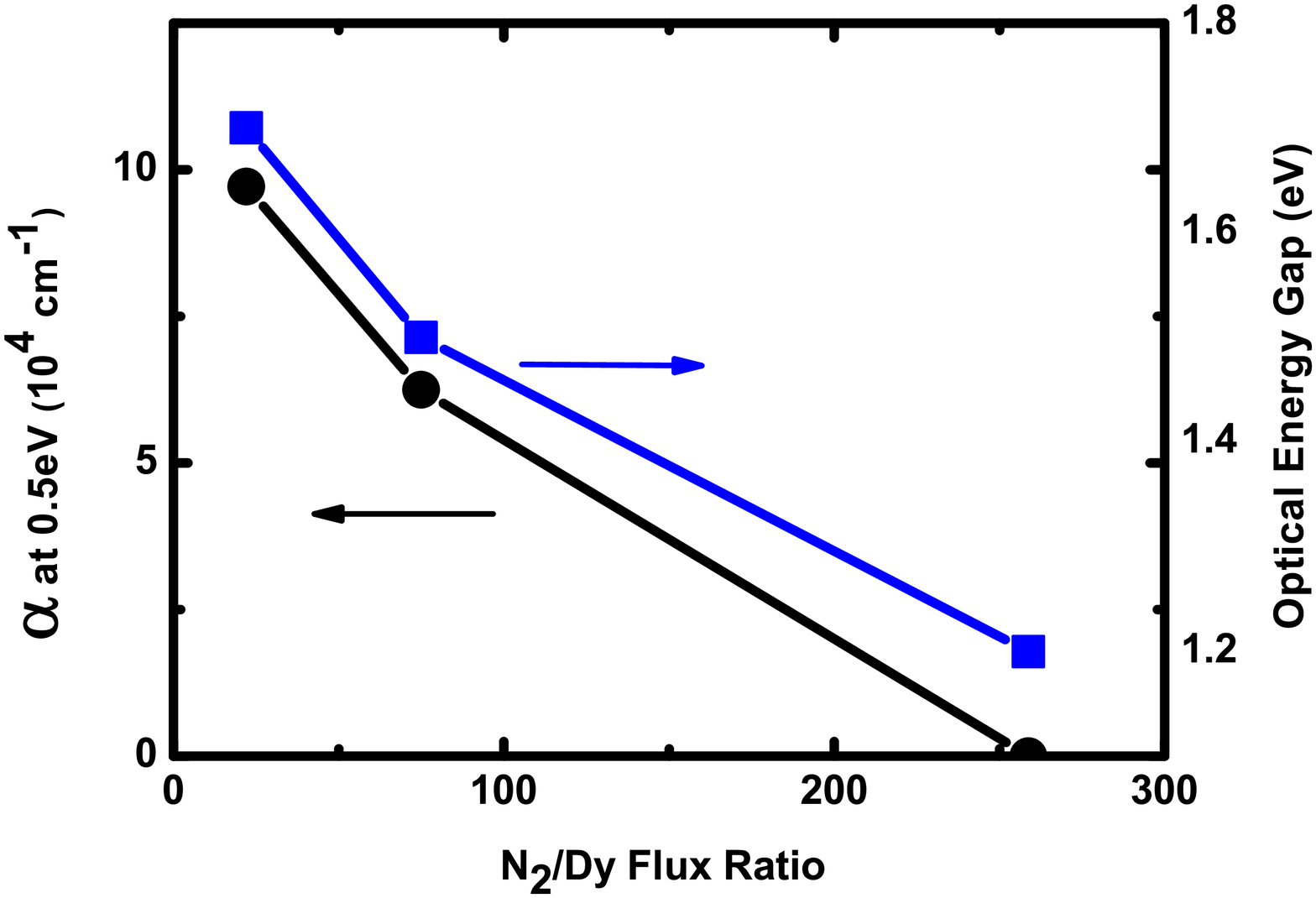}
\caption{(Color online)  Free carrier absorption (black solid circles) and the optical gap (blue solid squares) vs. N$_2$/Dy flux ratio during growth. Films grown with low N$_2$/Dy ratios show enahnced free carrier absorption and a significant Moss-Burstein shift.  }
\end{figure}

Turning now to the far infrared data we show in Figure 6 the reflectivity of the bare YSZ substrate and of the Si-capped DyN film on the YSZ substrate. These data can be fitted with only two damped oscillators; the dominant TO phonon expected in the NaCl structure is here at 280 cm$^{-1}$, damping constant 160 cm$^{-1}$; the frequency is somewhat lower than the estimated 338 cm$^{-1}$  based on an LSDA+$U$ approximation.\cite{granville09} The mode gives a contribution of 20 to the dc dielectric constant (Figure 7), though this number is sensitive to the assumed film thickness, which was not accurately known. The satisfactory fit required also a weaker resonance at 1200 cm$^{-1}$, damping 2400 cm$^{-1}$ and dielectric contribution of 1.8.  We assign this to a transition from nitrogen vacancy states expected to lie close below the conduction band.\cite {punya11} The fit also returns a high-frequency dielectric constant of 4.4, in reasonable agreement with the near IR refractive index of 2.0$\pm$0.1. 

\begin{figure} [h]
\centering
\includegraphics [scale=0.50]{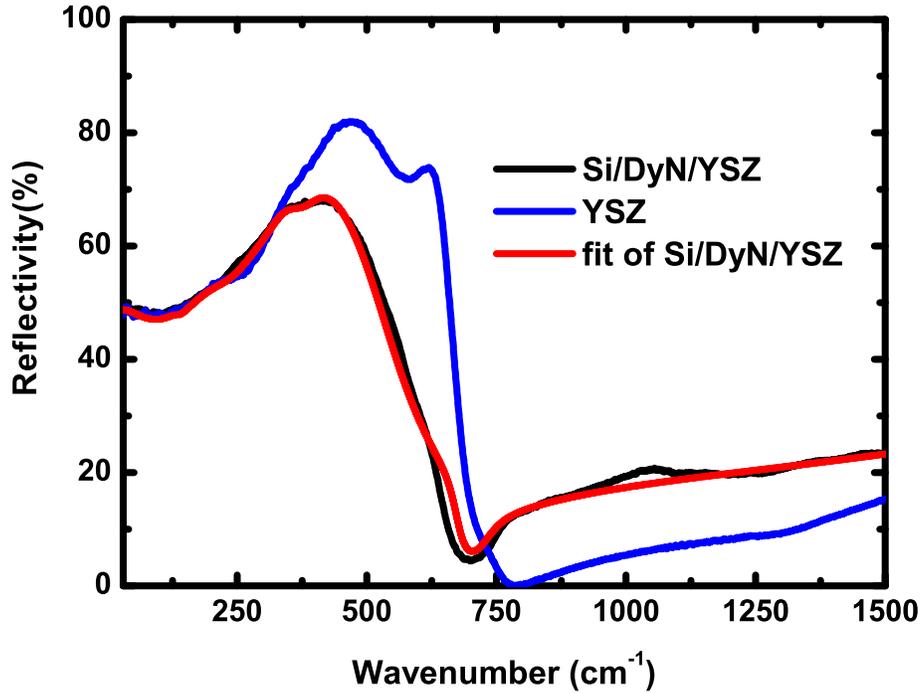}
\caption{(Color online) Infrared reflectivity spectrum of a Y-stabilized ZrO$_2$ substrate, a DyN thin film on a YSZ substrate capped by amorphous Si, and the fit to the data.}
\end{figure}

\begin{figure} [h]
\centering
\includegraphics [scale=0.50]{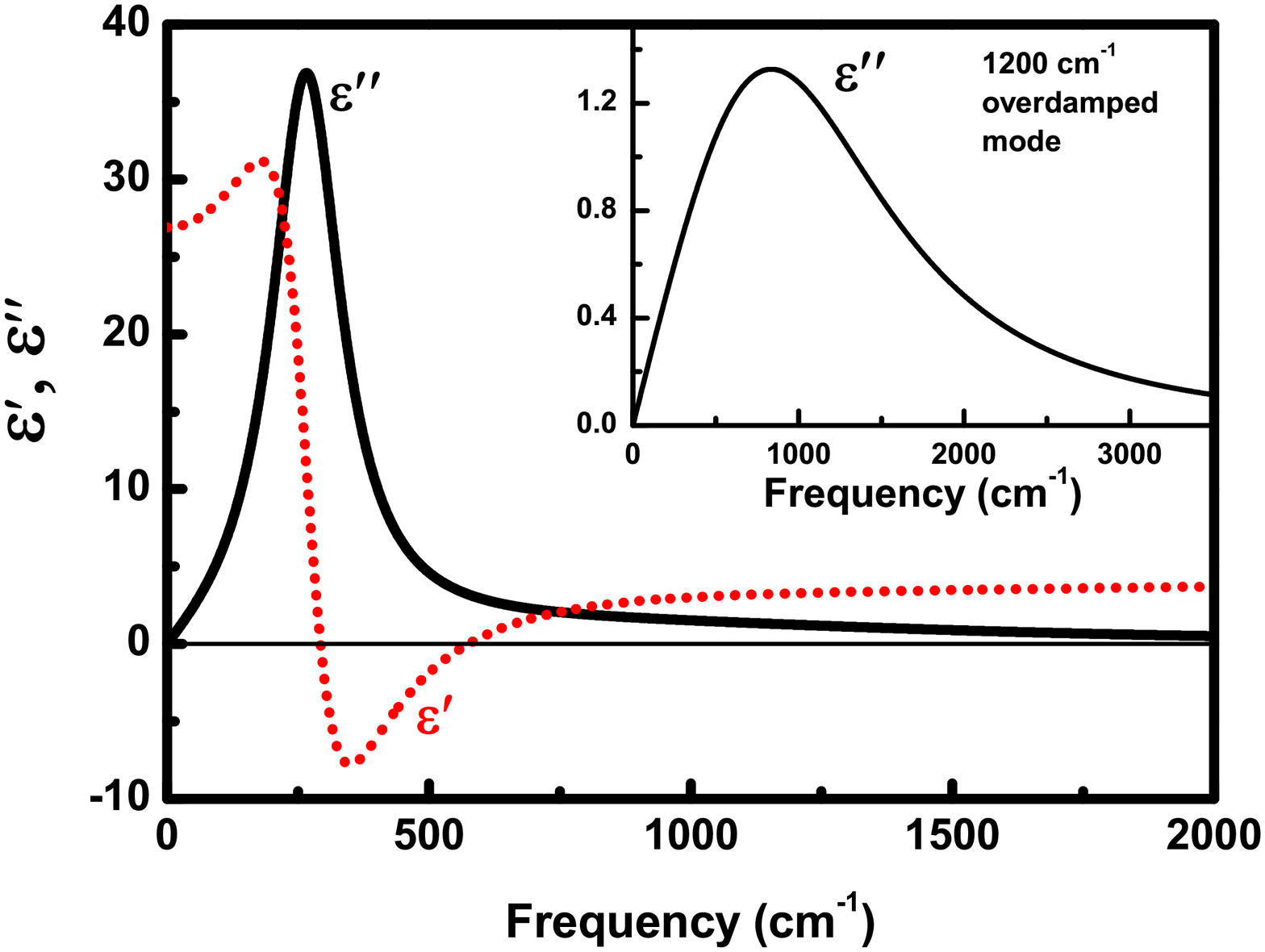}
\caption{ (Color online) Real and imagniary parts of complex dielectric function showing the polar phonon and nitrogen-vacancy donor to conduction band transition. }
\end{figure}

\section {Summary}

The optical response of DyN has been measured from 0.005 to 5.5 eV, covering both the lattice vibrational and interband regions. The direct interband gap is found at 1.2 eV in a near-stoichiometric film, with the absence of a measurable absorption below the gap establishing that DyN is a semiconductor. Films grown with sub-stoichiometric N concentration show free-carrier absorption below the gap, along with a blue-shifted absorption edge that is associated with the Moss-Burstein effect.  The excess absorption and the blue shift are a result of electrons released into the conduction band (CB) by nitrogen vacancies. The refractive index is 2.0$\pm$0.1. Far IR results show a value of 4.4 for the high frequency dielectric function, in good agreement with the near IR refractive index. The TO phonon has a frequency of 280 cm$^{-1}$, close to the value predicted by an LSDU+$U$ treatment. There is also evidence in the far IR data for a nitrogen-vacancy donor to conduction band transition at 1200 cm$^{-1}$.

\begin{acknowledgments}

This work was supported by MacDiarmid Institute for Advanced Materials and Nanotechnology, funded by the New Zealand Centres of Excellence Fund, NZ FRST (Grant No. VICX0808), the Marsden Fund (Grant No. 08-VUW-030) and the Czech Science Foundation (Projects No. P204/12/1163).

\end{acknowledgments}

\






\begin{thebibliography}{99}

\bibitem{Ruck08}
B. J. Ruck, Nanomagnetism and Spintronics (World Scientific, 2008), pp. 193-221.

\bibitem{Ivanov04}
V. A. Ivanov, T. G. Aminov, V. M. Novotortsev and V. T. Kalinnikov, Russ. Chem. Bull. \textbf{53}, 2357 (2004).

\bibitem{larson07}
P. Larson, W. R. L. Lambrecht, A. Chantis and M. van Schilfgaarde, Phys. Rev. B \textbf{75}, 045114 (2007).

\bibitem{aerts04}
C. M. Aerts, P. Strange, M. Horne, W. M. Temmerman, Z. Szotek and A. Svane, Phys. Rev. B \textbf{69}, 045115 (2004).

\bibitem{granville09}
S. Granville, C. Meyer, A. R. H. Preston, B. M. Ludbrook, B. J. Ruck, H. J. Trodahl, T. R. Paudel and W. R. L. Lambrecht, Phys. Rev. B \textbf{79}, 054301 (2009).

\bibitem{granville06}
S. Granville, B. J. Ruck, F. Budde, A. Koo, D. J. Pringle, F. Kuchler, A. R. H. Preston, D. H. Housden, N. Lund, A. Bittar, G. V. M. Williams and H. J. Trodahl, Phys. Rev. B \textbf{73}, 235335 (2006).


\bibitem{leu05}
F. Leuenberger, A. Parge, W. Felsch, K. Fauth and M. Hessler, Phys. Rev. B \textbf{72}, 014427 (2005).

\bibitem{khazen06}
K. Khazen, H. J. von Bardeleben, J. L. Cantin, A. Bittar, S. Granville, H. J. Trodahl and B. J. Ruck, Phys. Rev. B \textbf{74}, 245330 (2006).

\bibitem{mitra08}
C. Mitra and W. R. L. Lambrecht, Phys. Rev. B \textbf{78}, 195203 (2008).

\bibitem{sharma10}
A. Sharma and W. Nolting, Phys. Rev. B \textbf{81}, 125303 (2010).

\bibitem{ludbrook09}
 M. Ludbrook, I. L. Farrell, M. Kuebel, B. J. Ruck, A. R. H.Preston, H. J. Trodahl, L. Ranno, R. J. Reeves, and S. M. Durbin, J. Appl. Phys. \textbf{106}, 063910 (2009).

\bibitem{yosh11}
H. Yoshitomi, S. Kitayama, T. Kita, O. Wada, M. Fujisawa, H. Ohta and T. Sakurai, Phy. Status Solidi \textbf{8}, 488 (2011).

\bibitem{yoshprb}
H. Yoshitomi, S. Kitayama, T. Kita, O. Wada, M. Fujisawa, H. Ohta and T. Sakurai, Phys. Rev. B \textbf{83}, 155202 (2011).

\bibitem{tro07}
H. J. Trodahl, A. R. H. Preston, J. Zhong, B. J. Ruck, N. M. Strickland, C. Mitra and W. R. L. Lambrecht, Phys. Rev. B \textbf{76}, 085211 (2007)

\bibitem{rich11}
J. H. Richter, B. J. Ruck, M. Simpson, F. Natali, N. O. V. Plank, M. Azeem, H. J. Trodahl, A. R. H. Preston, B. Chen, J. McNulty, K. E. Smith, A. Tadich, B. Cowie, A. Svane, M. van Schilfgaarde and W. R. L. Lambrecht, Phys. Rev. B  \textbf{84}, 235120 (2011).

\bibitem{preston07}
A. R. H. Preston, S. Granville, D. H. Housden, B. Ludbrook, B. J. Ruck, H. J. Trodahl, A. Bittar, G. V. M. Williams, J. E. Downes, A. DeMasi, Y. Zhang, K. E. Smith and W. R. L. Lambrecht, Phys. Rev. B  \textbf{76}, 245120 (2007).

\bibitem{pokrant01}
S. Pokrant and J. A. Becker, Eur. Phys. J. D \textbf{16}, 165 (2001).

\bibitem{sclar64}
N. Sclar, J. Appl. Phys. \textbf{35}, 5 (1964).

\bibitem{bush69}
G. Busch and P. Wachter, Helv. Phys. Acta \textbf{42}, 930 (1969).


\bibitem{dismukes70}
J. P. Dismukes, W. M. Yim, J. J. Tietjen and R. E. Novak, J. Cryst. Growth \textbf{9}, 295 (1971).

\bibitem{punya11}
A. Punya, T. Cheiwchanchamnangij, and W. R. L. Lambrecht, Mater. Res. Soc. Symp. Proc. Vol. 1290 (2011)

\bibitem{zele98}
V. \u{Z}elezn\'{y}, I. Fedorov and J. Petzelt, Czech. J. Phys. \textbf{48}, 537 (1998).
\end{thebibliography}

\end{document}